\documentclass[twocolumn,epjc3]{svjour3}

\usepackage{t1enc}
\usepackage[latin2,utf8]{inputenc}

\usepackage{graphicx}
\graphicspath{ {./figs/} }

\RequirePackage{graphicx}
\RequirePackage{mathptmx}      
\RequirePackage{flushend}
\RequirePackage[numbers,sort&compress]{natbib}
\RequirePackage[colorlinks,citecolor=blue,urlcolor=blue,linkcolor=blue]{hyperref}

\RequirePackage{amsfonts}
\RequirePackage{amsmath,amssymb}
\RequirePackage{bm}

\usepackage[utf8]{inputenc}

\usepackage[normalem]{ulem}

\begin{document}

\titlerunning{The ReBB model and its $H(x)$ scaling version at 8 TeV: Odderon exchange is a certainty} 
\authorrunning{ I. Szanyi, T. Cs\"{o}rg\H{o}}
\title{The ReBB model and its $H(x)$ scaling version at 8 TeV: Odderon exchange is a certainty}

\author{
I. Szanyi\, \thanksref{e5,addr1,addr2,addr3}
\and
T. Cs\"{o}rg\H{o}\, \thanksref{e1,addr2,addr3}
}

\thankstext{e5}{e-mail: istvan.szanyi@cern.ch}
\thankstext{e1}{e-mail: tcsorgo@cern.ch}

\institute{
          E\"otv\"os University, H - 1117 Budapest, P\'azm\'any P. s. 1/A, Hungary\label{addr1}
          \and
Wigner FK, H-1525 Budapest 114, POB 49, Hungary\label{addr2}
          \and
          MATE Institute of Technology,  K\'aroly R\'obert Campus, H-3200 Gy\"ongy\"os, M\'atrai \'ut 36, Hungary\label{addr3}
	      }

\maketitle

\begin{abstract}
The Real Extended Bialas-Bzdak (ReBB) model is shown here to describe, in the $0.37 \le -t \le 1.2$ GeV$\null^2$ region, the proton-proton elastic differential cross section data published by the TOTEM Collaboration at LHC at $\sqrt{s} = 8 $ TeV center of mass energy.  In this kinematic range,  corresponding to the diffractive minimum-maximum region, a model-dependent Odderon signal higher than 18 $\sigma$ is obtained by comparing the ReBB model prediction for the $p\bar{p}$ elastic differential cross section to this TOTEM measured $pp$ elastic differential cross section data at 8 TeV. However, when combining this signal with the Odderon signals from the ReBB model in the $0.37 \leq -t \leq 1.2$ GeV$\null^2$ four-momentum-transfer range at $\sqrt{s} = $ 1.96, 2.76 and 7 TeV, it turns out that the combined significance is dominated not by the new 8 TeV but by that of earlier 7 TeV TOTEM data, that carry an even larger Odderon effect. Thus, in any practical terms, within the framework of the ReBB model, the Odderon signal in the limited $0.37 \leq -t \leq 1.2$ GeV$\null^2$ and $1.96 \leq \sqrt{s} \leq 8$ TeV kinematic region is not a probability, but a certainty. We show also that the $H(x)$ scaling version of the ReBB model works reasonably well at 8 TeV in the $0.37 \le -t \le 0.97$ GeV$\null^2$ region.
\end{abstract}

\section{Introduction}

In a recent paper \cite{Csorgo:2020wmw}, published in July 2021, we showed that the Real Extended $p=(q,d)$ version of the Bialas-Bzdak (ReBB) model developed in Ref.~\cite{Nemes:2015iia} based on the original papers, Refs.~\cite{Bialas:2006kw,Bialas:2006qf}, and later improvements, Refs.~\cite{Nemes:2012cp,CsorgO:2013kua}, describes in a statistically acceptable manner the proton-proton ($pp$) and proton-antiproton ($p\bar{p}$) scattering data in the kinematic range of $0.546\leq\sqrt{s}\leq 7$ TeV and $0.37\leq -t\leq1.2$ GeV$^2$. This model reproduces, in a statistically not excludable manner, both the observed minimum-maximum structure in the $pp$ differential cross section and the shoulder structure in $p\bar{p}$ differential cross section up to $\sqrt{s} = 7 $ TeV~\cite{Csorgo:2020wmw}. With these results at hand, we reported an at least 7.08 $\sigma$,  discovery level Odderon effect by comparing the $pp$ and $p\bar{p}$ differential cross sections at the same energies utilizing model-dependent extrapolations of the differential cross-sections of elastic $pp$ scattering to $\sqrt{s}$ $=$ 1.96 TeV and elastic $p\bar p$ scattering up to the lowest measured energy at LHC, 2.76 TeV. In that work, we could not evaluate the statistical significance of the Odderon signal at 7 TeV, because this signal was too large (greater than 10 $\sigma$ significance) to be quantified with the usual statistical and mathematical tools of CERN software package Root or with other software packages like MS Excel or Wolfram Mathematica.

We emphasize that the ReBB model is based on R. J. Glauber’s multiple diffraction theory, so it operates directly on the level of the elastic scattering amplitude of $pp$ and $p\bar p$ collisions. Thus we do not start with assuming a C-odd and C-even amplitude component in the ReBB model, in contrast to typical calculations based on Regge phenomenology. Instead,  we start with the evaluation of the $pp$ and the $p\bar p$ amplitudes from Glauber's multiple diffraction theory and fit the parameters of the resulting differential cross-sections to experimental data directly.  Finally we obtain the C-even (Pomeron) and C-odd (Odderon) components of the elastic scattering amplitude  as the average and the difference of elastic proton-antiproton and proton-proton amplitudes.  For the details of this kind of calculations, see for example Appendix C of our earlier paper, Ref.~\cite{Csorgo:2020wmw}.

In February 2021,  we and our colleagues published a model-independent Odderon effect with a significance of 6.26 $\sigma$ in Ref.~\cite{Csorgo:2019ewn}.
With the help of a previously unconsidered $H(x,s)$ scaling function, we explored the scaling properties of the differential cross-sections of the elastic $pp$ and $p\bar{p}$  collisions in a limited TeV energy range. By introducing this $H(x,s)$ scaling function, we scaled out from the measured elastic differential cross-sections the measurables that characterize the elastic scattering at $t=0$, \textit{i.e.}, the nuclear slope $B(s)$,
the real to imaginary ratio $\rho_0(s)$, the elastic cross-section $\sigma_{el}(s)$ and the total cross-section $\sigma_{tot}(s)$. Directly comparing the $H(x,s)$ scaling functions from measured TOTEM  $pp$ differential cross-section data at $\sqrt{s} = 2.76$ and 7 TeV, we observed that within errors these functions are energy independent in this TeV range. We could thus compare these
$pp$ scaling functions to that of the elastic $p\bar p$ differential cross-section data measured by the D0 collaboration at $\sqrt{s} = 1.96$ TeV. Our results thus provided a model-independent evidence for a $t$-channel Odderon exchange at TeV energies, with a significance of at least 6.26 $\sigma$, however, there was a model-dependent limitation in establishing the domain of validity of the energy independence of  $H(x,s|pp) = H(x, s_0|pp)$. We concluded that work~\cite{Csorgo:2019ewn} with a model-dependent evaluation of the domain of validity of the new scaling and its violations, utilizing the Real-Extended Bialas-Bzdak model~\cite{Csorgo:2020wmw,Nemes:2015iia}.
Using this model, we have shown, that the $H(x,s|pp)$ scaling function for \break elastic $pp$ collisions becomes energy independent,\break $H(x,s|pp) = H(x,s_0|pp)$, within the center of mass energy range of   $200~\hbox{GeV}\leq\sqrt{s}\leq7$~TeV,~with a $-t$ range gradually narrowing with decreasing center of mass energies.
Based on a direct comparison of experimental data
on $H(x,s_0|pp)$ and $H(x,s|pp)$, we observed 
that in the TeV energy range, the elastic proton-proton differential cross-sections obey an energy independent, data collapsing behaviour~\cite{Csorgo:2019ewn}. 

We have also participated in the joint paper of the  TOTEM and D0 Collaborations~\cite{TOTEM:2020zzr}, published in August 2021, that showed an {\it almost} model-independent 
Odderon effect with a statistical significance of 5.2 $\sigma$. This significance was obtained by combining two kind of Odderon signals, $\sigma_1$ and $\sigma_2$.
A statistical significance of $\sigma_1 \ge 3.4$ $\sigma$ was obtained from the comparison of D0 measured 
elastic $p\bar p$ scattering data in the diffractive shoulder region
with D0-TOTEM extrapolated $pp$ data evaluated at $\sqrt{s} = 1.96$ TeV. Combining $\sigma_1$ with $\sigma_2$, the result from the comparison of TOTEM measurements of the real-to-imaginary ratio $\rho_0(s)$ and total cross section at $\sqrt{s} = 13$ TeV
with a set of models assuming that the considered set of models is a representative, characteristic sample from among the possible models, give a combined significance ranging from 5.2 to 5.7 $\sigma$.

The assumption that the model class considered by TOTEM at $t= 0$ and $\sqrt{s} = 13$ TeV in Refs.~\cite{TOTEM:2017sdy,TOTEM:2020zzr} 
is a representative sample of models 
has been questioned in Refs.~\cite{Donnachie:2019ciz,Donnachie:2022aiq}: Donnanchie and Landshoff presented a Regge-motivated model, that explains the low $-t$ data of TOTEM at $\sqrt{s} = 13$ TeV without an Odderon contribution, suggesting that the statistical significance of the Odderon contribution $\sigma_2$ is practically zero for at least one of the reasonably possible models.
Donnachie and Landshoff argue, that there is
a good reason to believe in an Odderon contribution at
large $-t$, which is identified with triple-gluon exchange effects in QCD, as predicted in Ref.~\cite{Donnachie:1983ff}. 
Indeed, an indication for such an effect at  3.35 $\sigma$
level has been seen by Breakstone et al, already at the ISR energy of $\sqrt{s} = 53$ GeV~\cite{Breakstone:1985pe}. However, this energy is not yet high enough to exclude possible Reggeon exchange effects: one needs to reach the TeV energy scale to make the Pomeron and possible Odderon contributions dominant over the exchanges of other hadronic resonances or Reggeon exchanges~\cite{Broniowski:2018xbg}.
Indeed, a statistically significant Odderon signal in the diffractive minimum-maximum region, with a statistical significance of at least 6.26 $\sigma$ has been reported recently by us and our collaborators in Ref.~\cite{Csorgo:2019ewn}. In this direct data-to-data comparison of the $H(x)$ scaling functions of elastic $pp$ and $p\bar p$ scattering, the results by definition are insensitive to $t=0$ observables, but are sensitive to differences of $pp$ and $p\bar p$
scattering in the diffractive minimum-maximum region.


In the present, relatively short and focused manuscript, we cannot review more detailed background of the discovery of Odderon.
Instead, we focus on the analysis of the  new data that were published recently by the TOTEM Collaboration on elastic $pp$ scattering in Ref.~\cite{TOTEM:2021imi}. In this manuscript, we show that the ReBB model --- whose fit parameter values were determined before in Ref.~\cite{Csorgo:2020wmw} --- describes in a statistically acceptable way these new and final, 8 TeV $pp$ differential cross section data measured by TOTEM ~\cite{TOTEM:2021imi} in the $-t$ range used for the calibration of the ReBB model,
namely the $0.37 \leq -t \leq 1.2 $ GeV$^2$ kinematic range, and results in an extremely large Odderon signal.
We show also that the $H(x)$ scaling limit of the ReBB model also describes the recent TOTEM measurements at 8 TeV,
however, in a more limited $-t$ range. 
We find, as detailed in the body of the manuscript, that the ReBB model results in an Odderon signal at $\sqrt{s} =$ 8 TeV, that
corresponds to a $CL=1-1.111\times 10^{-74}$ Odderon observation probability and an at least 18 $\sigma$ statistical significance, which is 
way above  the usual 5 $\sigma$ discovery threshold level. It turns out that standard numerical packages like Root, Mathematica, or Excel cannot evaluate precisely confidence levels and statistical significances at such extremely large significances: packages start to become unreliable typically  above the 8-10 $\sigma$ significance levels. Hence, we have developed some new analytic approximation schemes, and have summarized them in  \ref{sec:app1}, to evaluate significances and confidence levels for such extreme values of significances. Previously, we have reported in Ref.~\cite{Csorgo:2019ewn} that within the framework of the ReBB model, the statistical significance of the Odderon observation is also extremely high, above 10 $\sigma$ at $\sqrt{s} = 7$ TeV, but we were not yet able to quantify more precisely those significances. With the new analytic approximation scheme detailed in ~\ref{sec:app1}, we also evaluate now the significance of Odderon observation at 7 TeV within the framework of the ReBB model, and compare and combine the statistical significances of all of our ReBB model results obtained at 1.96, 2.76, 7, and 8 TeV.

This manuscript is organized as follows. In Sec.~\ref{sec:rebb_8_tev} we detail the description of the TOTEM 8 TeV $d\sigma/dt$ data by the ReBB model and discuss an observable model-dependent Odderon signal at 8 TeV. In Sec.~\ref{sec:hx_8_tev} we present the results concerning the description of the TOTEM 8 TeV $d\sigma/dt$ data by the \textit{H(x)} scaling version of the ReBB model. The results are discussed in detail in Sec.~\ref{sec:disc}. Finally, we summarize and conclude in Sec.~\ref{sec:summ}.

\section{ReBB model at 8 TeV}\label{sec:rebb_8_tev}

The $p=(q,d)$ version of the ReBB model \cite{Nemes:2015iia} describes the proton as a bound state of a constituent quark and a diquark. The free parameters of the model are the Gaussian radii of the quark, the diquark, and the separation between them, correspondingly, $R_q$, $R_d$, and $R_{qd}$ and also an opacity parameter denoted by $\alpha$. The model contains two additional fit parameters: the ratio of the quark and diquark masses, $\lambda$, and the normalisation parameter appearing in the inelastic quark-quark cross section, $A_{qq}$. It was shown in Ref.~\cite{Nemes:2012cp} and later confirmed in Ref.~\cite{Csorgo:2020wmw} that $A_{qq}$ can be fixed at a value of 1.0 while $\lambda$ can be fixed at a value of 0.5.

In Ref.~\cite{Csorgo:2020wmw} we determined the energy dependence of the ReBB model parameters, $R_q$, $R_d$, $R_{qd}$, and $\alpha$. We found that the energy dependence of the radius parameters are the same for $pp$ and $p\bar{p}$ scattering while the energy dependence of the opacity parameter $\alpha$ is different for the $pp$ and $p\bar{p}$ processes $i.e.$ there are different, $\alpha^{pp}$ and $\alpha^{p\bar{p}}$ parameters. The energy dependencies of all the five parameters in the energy range of $0.546 \leq \sqrt{s} \leq 7 $ TeV are consistent with a linear logarithmic \textit{i.e.} $p_0+p_1\ln (s/s_0)$ shape where $p_0$ and $p_1$ are fit parameters and $s_0=1$ GeV$^2$. Thus in the framework of the ReBB model in this energy range the $pp$ and $p\bar{p}$ data can be described by altogether $5\times 2=$ 10 parameters. This trend was verified in Ref.~\cite{Csorgo:2020wmw} by testing that the $\sqrt{s}$ dependent model parameters, taken from the best fits to  $p_0+p_1\ln (s/s_0)$ shapes indeed reproduce the available data sets at the TeV energy scale, in the $-t$ range of the calibration of these model parameters. This way it was determined that the ReBB model  gives a statistically acceptable, $CL \geq 0.1 \%$ level description of all available $pp$ and $p\bar p$
elastic scattering data in the $0.37 \leq -t \leq 1.2$ GeV$^2$ four-momentum and in the $0.546 \leq \sqrt{s} \leq 7$ TeV center of mass energy range.
At that time {\it preliminary} $pp$ data were also available from the TOTEM Collaboration at $\sqrt{s} = 8$ TeV. As indicated in Ref.~\cite{Csorgo:2020wmw}, these TOTEM preliminary data were not inconsistent with the ReBB model. 

\begin{figure}[!hbt]
 \centerline{
 \includegraphics[width=0.45\textwidth]{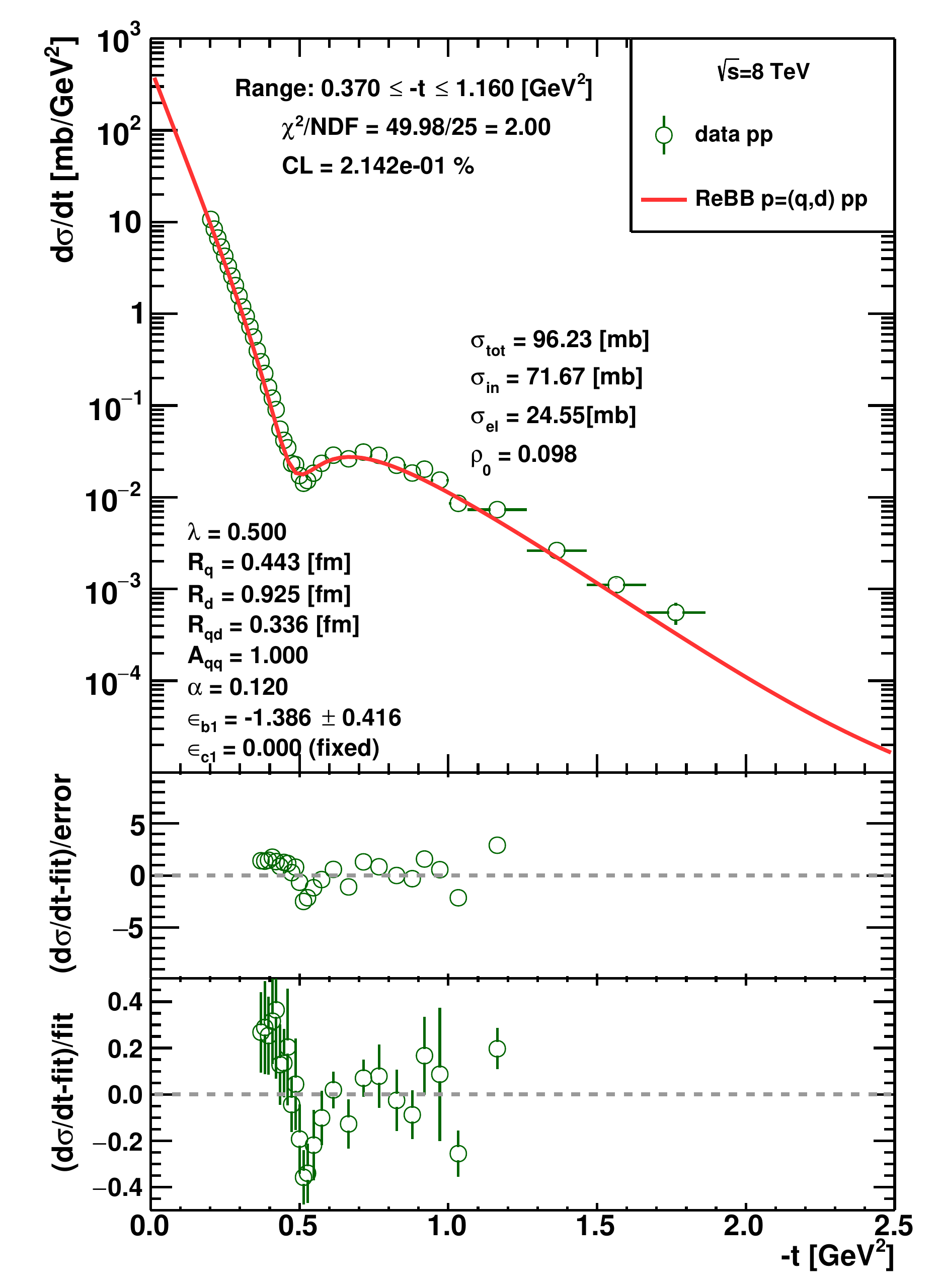}
 }
\caption{ 
Comparison of the $pp$ differential cross section calculated from the ReBB model --- using the energy calibration of the fit parameters done in Ref.~\cite{Csorgo:2020wmw} --- with the 8 TeV $pp$ differential cross section data measured and published by TOTEM in Ref.~\cite{TOTEM:2021imi}. According to this publication~\cite{TOTEM:2021imi}, we consider half of the $-t$ bin size of the data points as horizontal type A errors, when applying the $\chi^2$ definition detailed in Ref.~\cite{Csorgo:2020wmw}. On this Figure, the shown  values refer to $pp$ elastic scattering, 
corresponding to the physical observables at $t=0$, and the ReBB model parameter values obtained from those trends, that were determined  previously in Ref. ~\cite{Csorgo:2020wmw}. 
}
\label{fig:H(x)_Odderon_1}
\end{figure}

As an extension to Ref.~\cite{Csorgo:2020wmw}, in Fig.~\ref{fig:H(x)_Odderon_1} we show
the comparison of the $pp$ differential cross section calculated from the ReBB model --- using the energy calibration of the fit parameters done in Ref.~\cite{Csorgo:2020wmw} --- with the {\it final} 8 TeV $pp$ differential cross section data measured by TOTEM and published recently in Ref.~\cite{TOTEM:2021imi}. One can see that the energy-calibrated model, in its validity range, $0.37\leq -t \leq 1.2$ GeV$^2$, describes the data in a statistically acceptable manner with a confidence level of 0.2 \%. The used $\chi^2$ definition with correlation parameters, $\epsilon_B$ and $\epsilon_C$, together with the classification of measurement errors are detailed in Ref.~\cite{Csorgo:2020wmw} and are now fully based on Ref.~\cite{Adare:2008cg}.

Considering these results one can conclude that the ReBB model can be used to describe the data at 8 TeV in a limited kinematic region. This limited kinematic region is suitable to perform an Odderon search and  to extract an Odderon signal from the 8 TeV $pp$ TOTEM $d\sigma/dt$ data: As detailed and utilized recently in Refs.~\cite{Csorgo:2019ewn,Csorgo:2020wmw, TOTEM:2020zzr} a possible difference between $pp$ and $p\bar{p}$ measurable quantities at the TeV energy scale theoretically can be attributed only to the effect of a $t$-channel C-odd Odderon exchange. 

\begin{figure}[hbt!]
 \centerline{
 \includegraphics[width=0.45\textwidth]{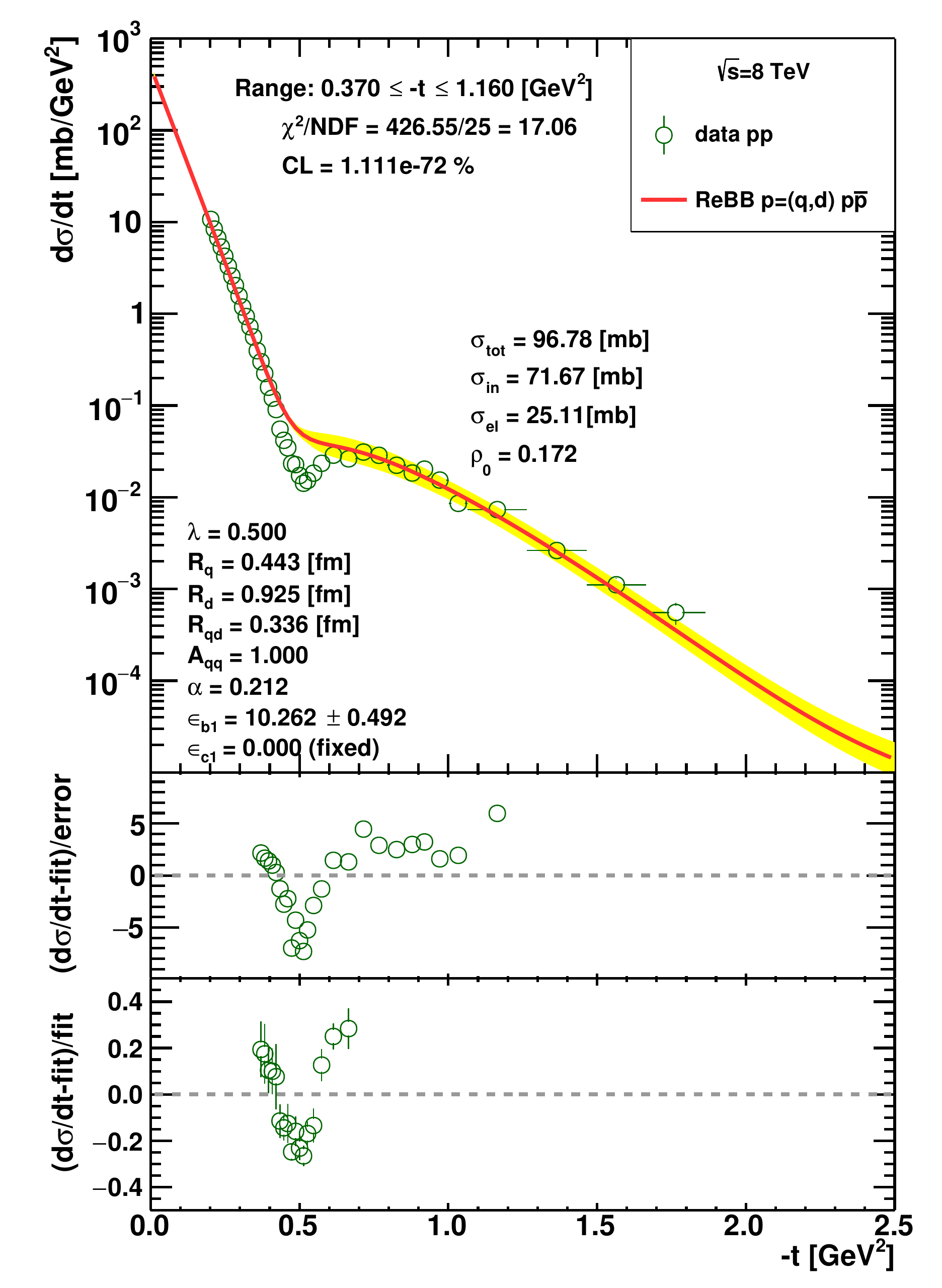}
 }
\caption{Comparison of the $p\bar{p}$ differential cross section calculated from the ReBB model --- using the energy calibration of the fit parameters done in Ref.~\cite{Csorgo:2020wmw} --- with the 8 TeV $pp$ differential cross section data measured by TOTEM \cite{TOTEM:2021imi}. 
Same as Fig.~\ref{fig:H(x)_Odderon_1}, but now the $p\bar p$ prediction of the ReBB model with its systematic error band.
The Odderon signal corresponds to the statistical difference between the yellow error-band, valid for $p\bar p$ ReBB model extrapolated data,
and the TOTEM measurements of $pp$ data at $\sqrt{s} = 8$ TeV.
On this Figure, the shown  values refer to $p\bar p$ elastic scattering, 
corresponding to the model predictions for $p\bar p$  physical observables at $t=0$ and the ReBB model parameter values for 
$p\bar p$ obtained from previously determined trends ~\cite{Csorgo:2020wmw}.
}
\label{fig:H(x)_Odderon_3}
\end{figure}

As an extension to the results obtained in Ref.~\cite{Csorgo:2020wmw}, we have compared the $p\bar{p}$ differential cross section calculated from the ReBB model --- using the energy calibration of the fit parameters done in Ref.~\cite{Csorgo:2020wmw} --- with the 8 TeV $pp$ differential cross section data measured by TOTEM \cite{TOTEM:2021imi}. This comparison  is shown in Fig.~\ref{fig:H(x)_Odderon_3}, which indicates a difference between the $pp$ and $p\bar{p}$ differential cross sections with a probability of essentially 1, corresponding to a $CL=1-1.111\times10^{-74}$, \textit{i.e.}, an Odderon observation with a statistical significance $\geq$18.28 $\sigma$. (For the details of the significance calculation, see \ref{sec:app1}.)

\begin{figure}[hbt!]
\centering
 \includegraphics[width=0.42\textwidth]{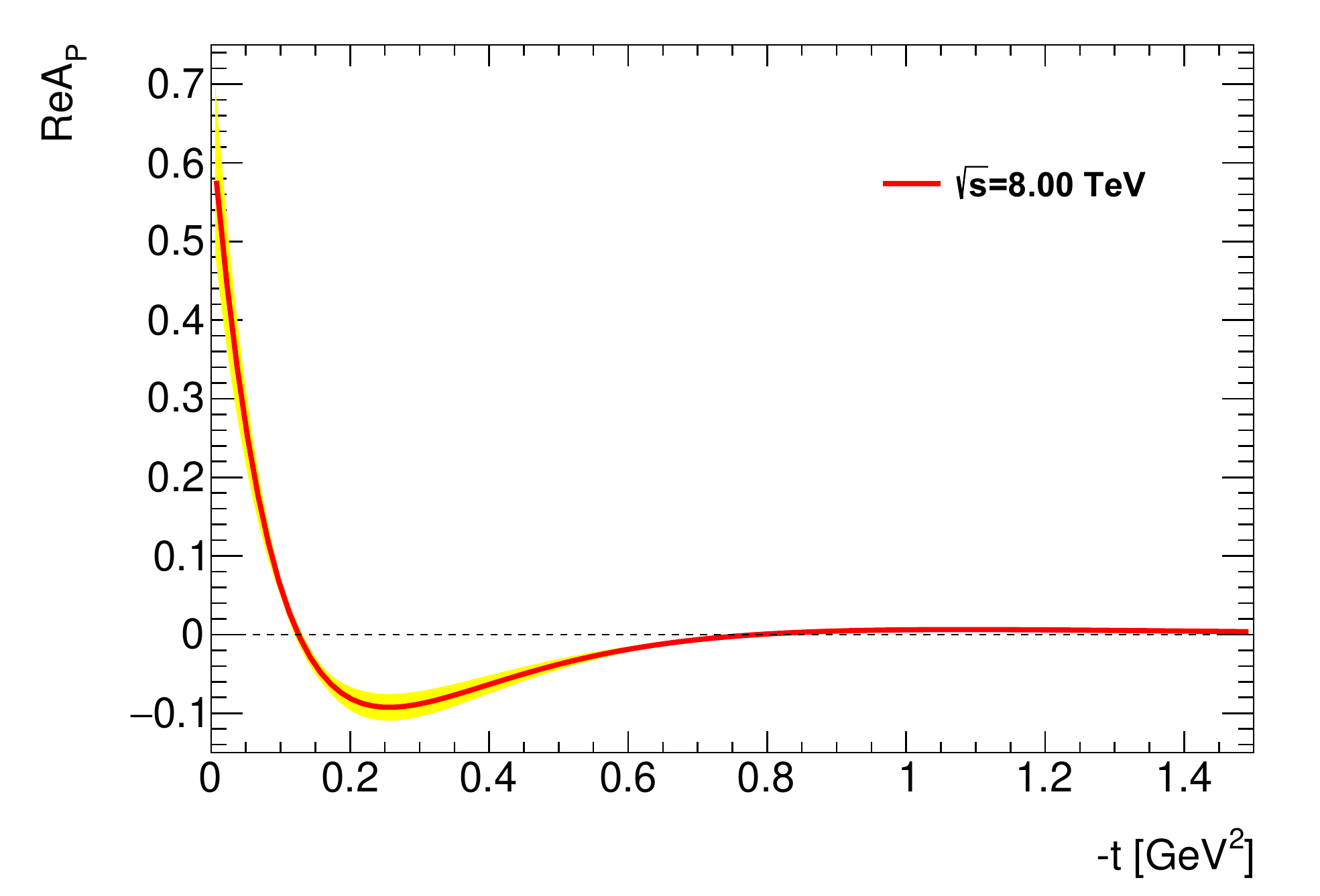}
  \includegraphics[width=0.42\textwidth]{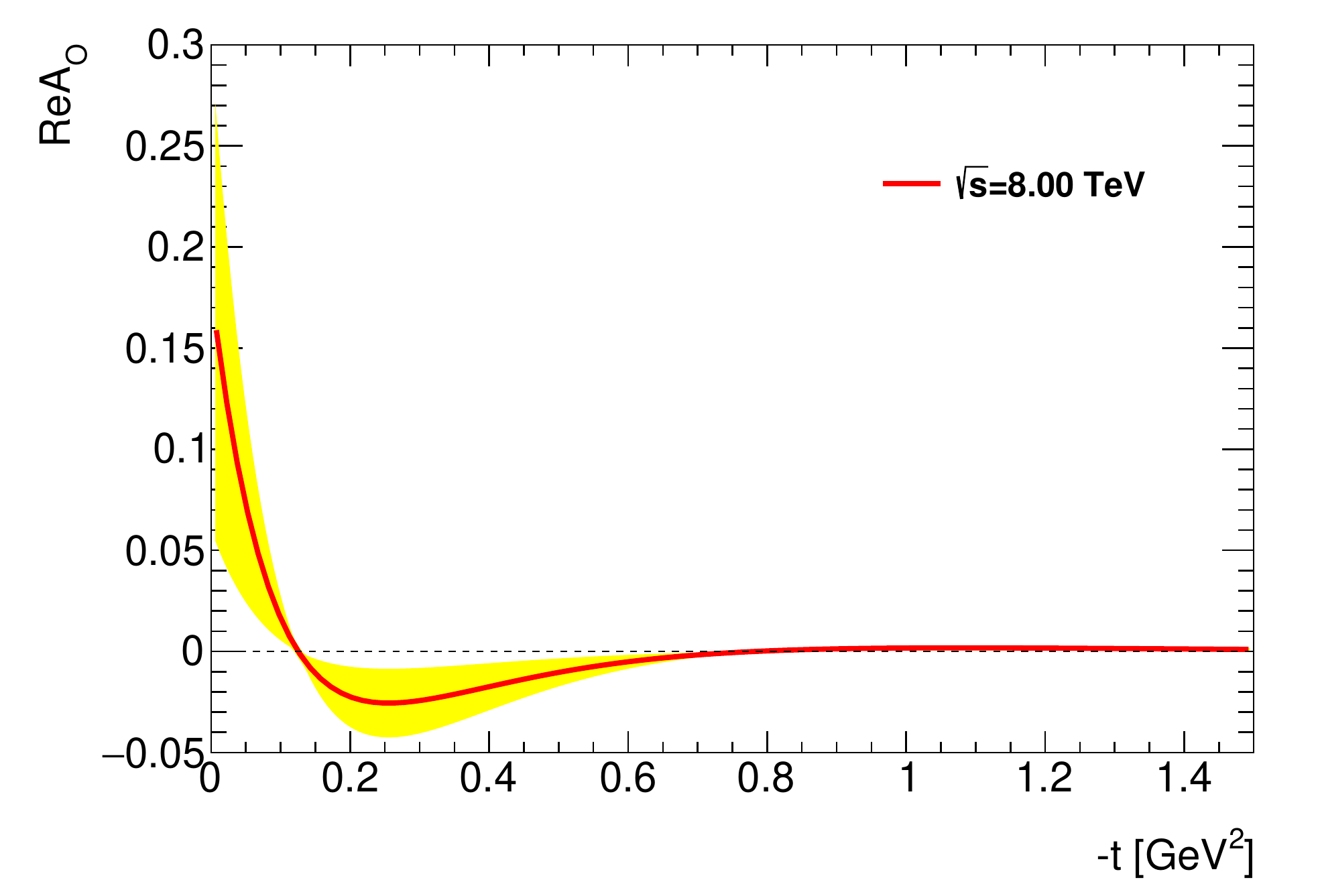}
   \includegraphics[width=0.42\textwidth]{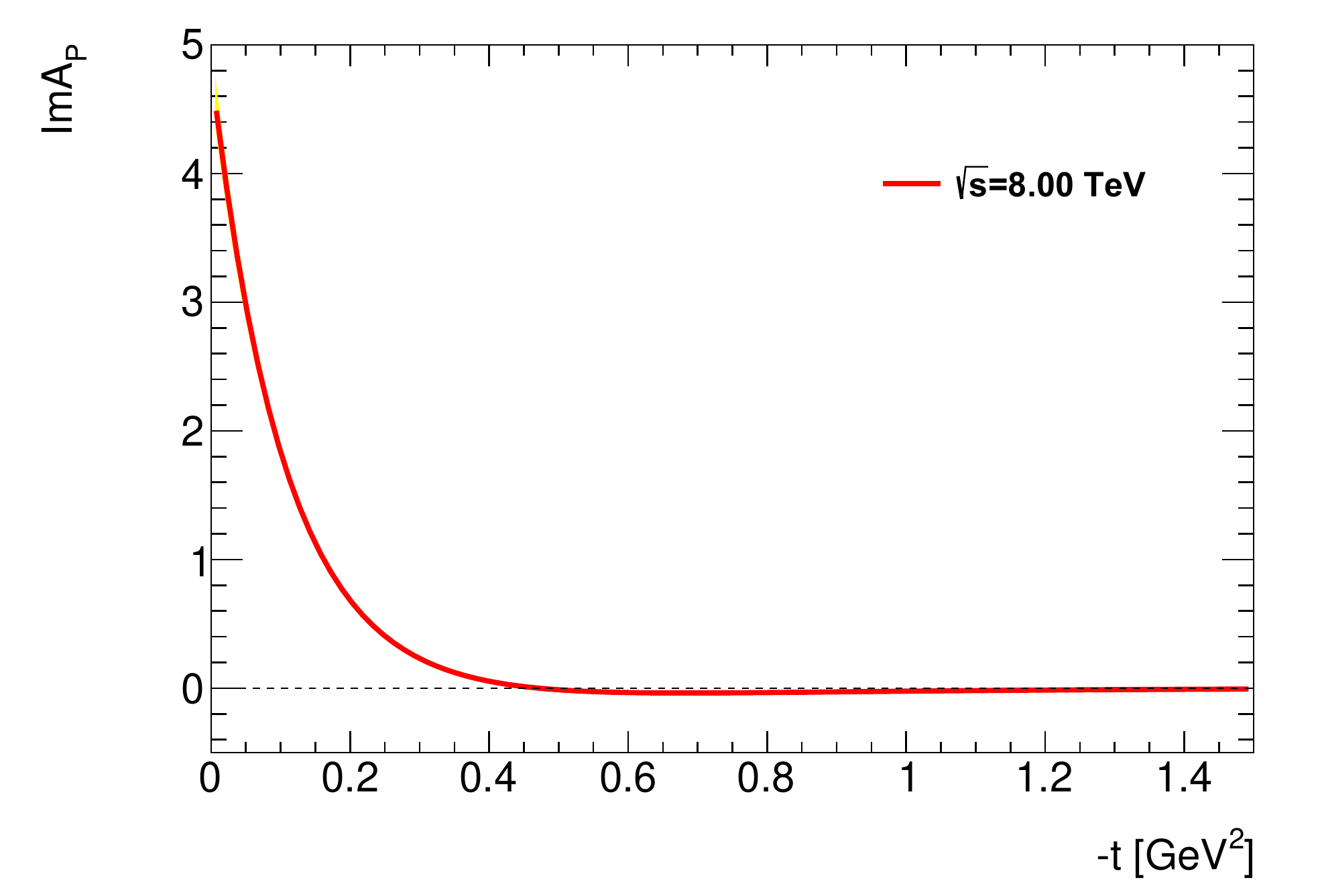}
    \includegraphics[width=0.42\textwidth]{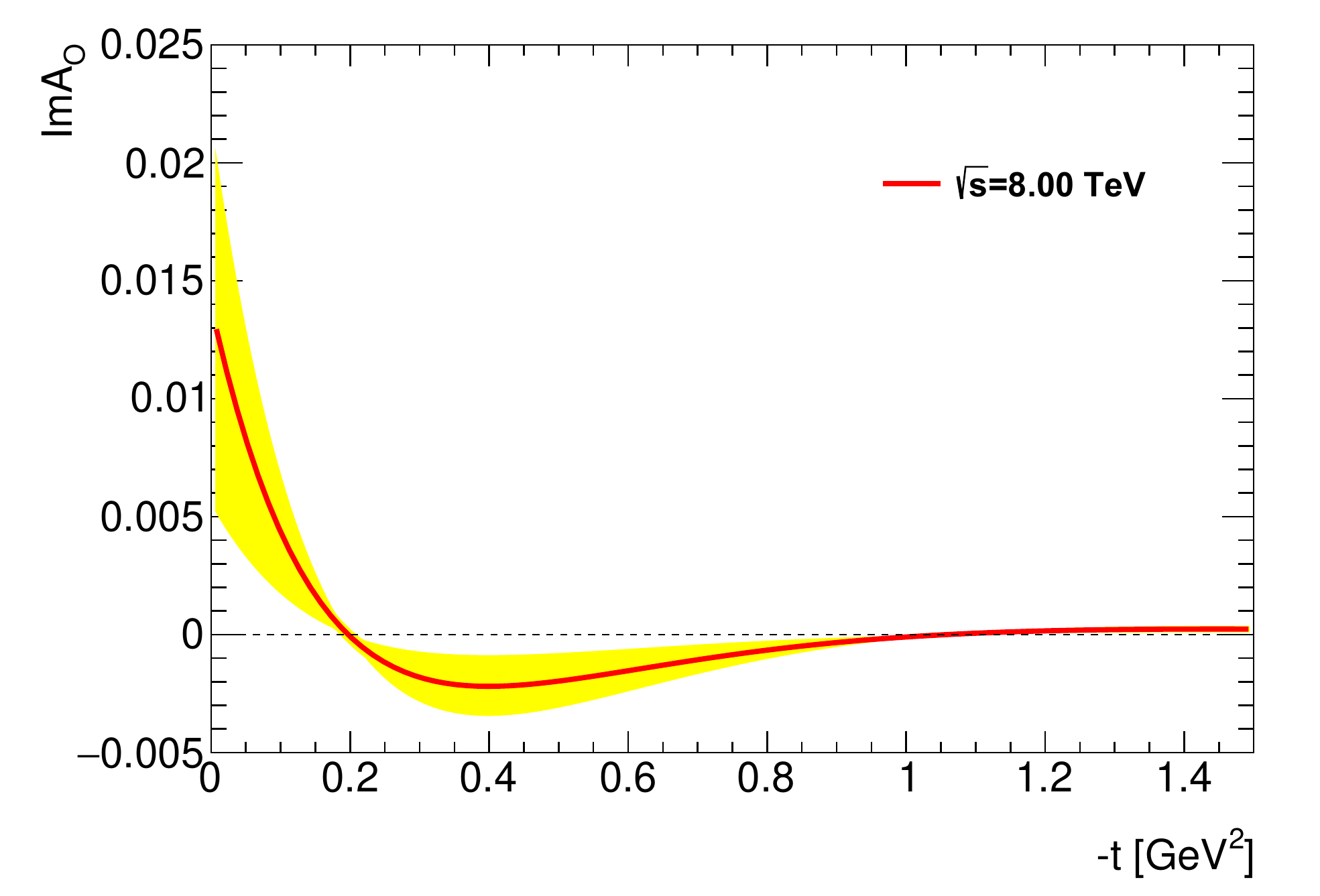}
\caption{Real and imaginary parts of the Pomeron (C-even) and Odderon (C-odd) components of the scattering amplitude at $\sqrt{s}=8$ TeV calculated from the ReBB model using the energy calibration of the fit parameters done in Ref.~\cite{Csorgo:2020wmw}. Yellow band indicates the systematic errors of the calculation. The ReBB model is validated in a limited $-t$ range, as detailed in Ref.~\cite{Csorgo:2020wmw}, hence we consider its predictions in the small-$|t|$ region on the qualitative level, only.}
\label{fig:ImRe_even_odd}
\end{figure}

Fig.~\ref{fig:ImRe_even_odd} shows the real and imaginary parts of the Pomeron (C-even) and Odderon (C-odd) components of the scattering amplitude at $\sqrt{s}=8$ TeV calculated from the ReBB model using the energy calibration of the fit parameters done in Ref.~\cite{Csorgo:2020wmw}. This amplitude results in the $pp$ and $p\bar{p}$ differential cross sections shown in Fig.~\ref{fig:H(x)_Odderon_1} and Fig.~\ref{fig:H(x)_Odderon_3}, respectively. One can see on Fig.~\ref{fig:ImRe_even_odd}, that at $t=0$ both the real and the imaginary parts are dominated by the Pomeron. Indeed, the real part of the Odderon is about an order of magnitude larger, than its imaginary part. The opposite is true for the Pomeron: its imaginary part is about an order of magnitude larger, than its real part. One can easily cross-check that the Odderon component of elastic proton-proton scattering at 8 TeV not only results in a difference in the dip region, but also results in a decrease of the real to imaginary ratio at $t=0$, as detailed in Fig. 26. of Ref.~\cite{Csorgo:2020wmw}.

\section{ReBB model at 8 TeV in the H(x) scaling limit}\label{sec:hx_8_tev}

The  scaling function $H(x,s)=\frac{1}{\sigma_{el} B_0}\frac{d\sigma}{dt}\big|_{x=-tB_0}$ was introduced in Ref.~\cite{Csorgo:2019ewn}, where $\sigma_{el}\equiv \sigma_{el}(s)$ is the elastic cross section, $B_0\equiv B(s,t\rightarrow0$) is the nuclear slope parameter, and $\frac{d\sigma}{dt}\equiv \frac{d\sigma}{dt}(s,t)$ is the elastic differential cross section. For elastic $pp$ scattering, 
$H(x,s) = H(x, s_0)$ was proven to be energy independent~\cite{Csorgo:2019ewn}  in an energy range that includes 1.96 TeV and 7 TeV and was used to extract an Odderon signal by data-data comparison. For a reference energy, $s_0 = 7$ TeV was chosen. 
If $H(x,s) = H(x, s_0) $ is energy-independent in a kinematic range of $s_1 \leq s \leq s_2 $ we say that the $H(x)$ scaling behavior is present in that $s_1 \leq s \leq s_2 $ range. 

The ReBB model manifests an $H(x)$ scaling limiting behavior if the following conditions are satisfied:
\begin{itemize}
    \item the energy dependencies of the radius parameters are determined in the $s_1 \leq s \leq s_2 $ energy range by the same $b(s)$ scaling function: $R_i(s) = b(s)R_{i0}$, where $i\in\{q,d,qd\}$ and $R_{i0}$ are the values of the physical scale parameters (Gaussian radii characterizing the size of the quark, diquark and their separation) at some reference energy $s_0$, \textit{i.e.}, $R_{i0}= R_i(s_0)$ with $b(s_0) = 1$;
    \item the opacity parameter, $\alpha$ is energy independent, $i.e.$, $\alpha(s)=\alpha(s_0)$ in the $s_1 \leq s \leq s_2 $ energy range.
\end{itemize}

\begin{figure}[!hbt]
\includegraphics[width=0.45\textwidth]{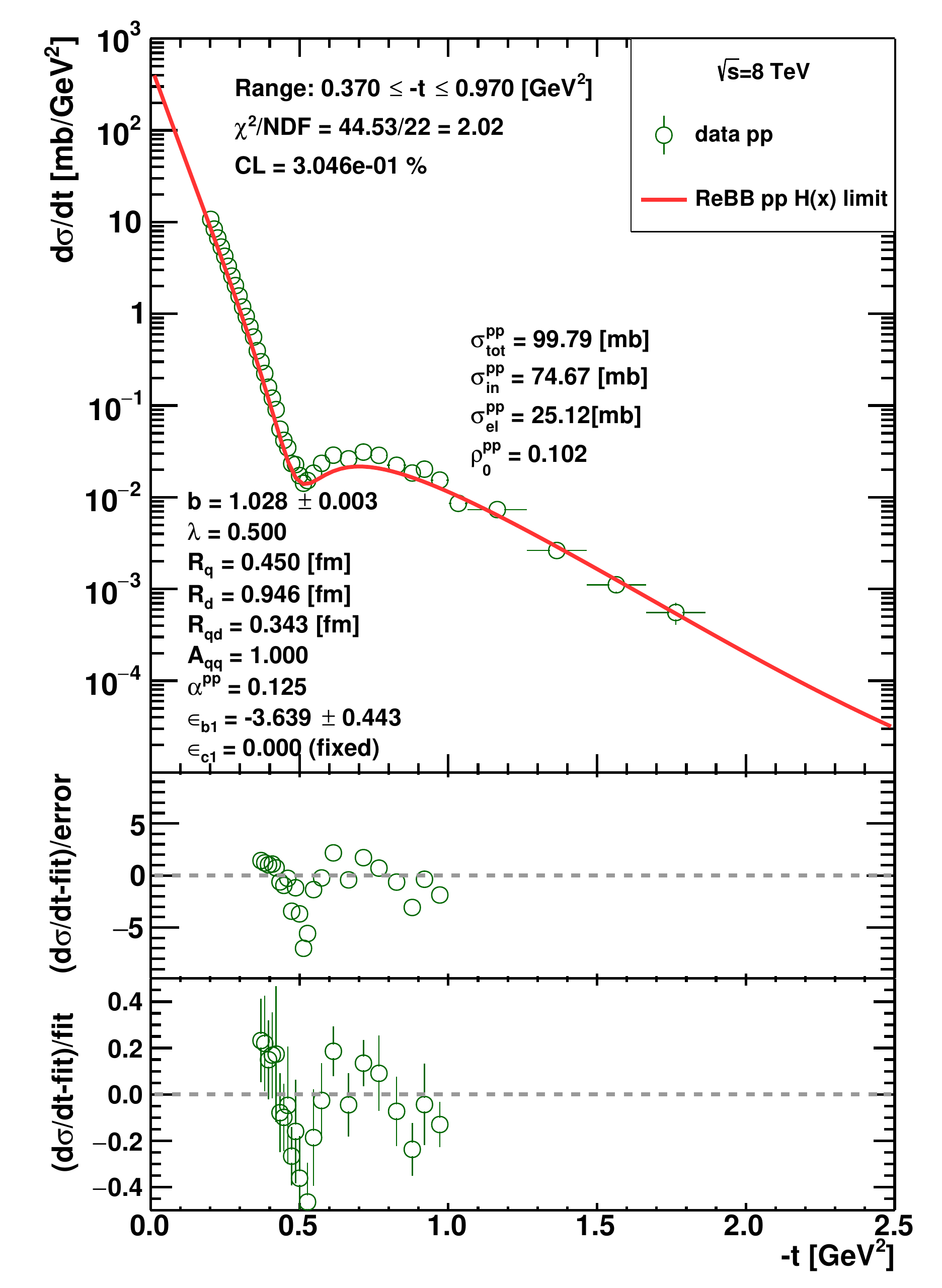}
\caption{The $H(x)$ scaling limit of the ReBB model compared to the $pp$ 8 TeV TOTEM data \cite{TOTEM:2021imi}. 
Otherwise, same as Fig.~\ref{fig:H(x)_Odderon_1}.
}
\label{fig:H(x)_Odderon_5}
\end{figure}

By choosing the reference energy to be $s_0=7$ TeV, and using the ReBB model parameter values obtained by fitting the TOTEM 7 TeV $d\sigma/dt$ data in Ref.~\cite{Csorgo:2020wmw} we performed a fit to the 8 TeV TOTEM $d\sigma/dt$ data by letting the value of the scaling function, $b(s)$ at $\sqrt{s}=8$ TeV to be free. The result is shown in Fig.~\ref{fig:H(x)_Odderon_5}. One can see that the $H(x)$ scaling version of the ReBB model describes the data at 8 TeV in a statistically acceptable way, with CL = 0.3 \%, in the $-t$ range of $0.37\leq-t\leq0.97$ GeV$^2$. This range is, however, somewhat narrower than the validity range of the original ReBB model which extends up to $-t=1.2$ GeV$^2$.  


\section{Discussion}\label{sec:disc}

As detailed in Section~\ref{sec:rebb_8_tev}, the ReBB model in its validity range in $-t$, $0.37\leq-t\leq1.2$ GeV$^2$, gives a statistically acceptable description to the 8 TeV $pp$ differential cross section data measured by the TOTEM Collaboration and predicts an Odderon observation at 8 TeV with an extremely high statistical significance $\geq$ 18.28 $\sigma$, corresponding to an Odderon observation that, in practical terms, has a probability essentially 1. In Section~\ref{sec:hx_8_tev}, we have shown also that the $H(x)$ version of the ReBB model works also reasonably well at 8 TeV, but only in a limited $-t$ range, extending only up to $0.97$ GeV$^2$.  In this section, we discuss and detail these results, and combine them with results obtained at other energies. 

First of all we need to clarify that in Ref.~\cite{Csorgo:2020wmw} we considered the $\epsilon_b$ and $\epsilon_c$ parameters of the non-traditional $\chi^2$ definition, originally introduced in Ref.~\cite{Adare:2008cg}, as free parameters, decreasing the number of degrees of freedom ($NDF$) as usual. However, we did not consider at that time that their central value of 0 and their error or variance of 1 is also a known information. However, after a careful study of Appendix A of Ref.~\cite{Adare:2008cg}, it turns out that the original version of this PHENIX method of diagonalizing the covariance matrix takes this information also into account: the $\epsilon_b$ and $\epsilon_c$ parameters should be considered not only as free parameters, increasing the number of fit parameters by 2, but also the extra information on their known central value and variance should be considered as two new data points or measurements. Thus the proper evaluation of the number of degrees of freedom is $NDF = (d+2) - (p+2) = d-p $, where $d$ stands for the number of measured data points and $p$ stands for the number of physical (model dependent) fit parameters. In Ref.
~\cite{Csorgo:2020wmw} we did not consider the information that increased the number of data points by the known central value and error on 
$\epsilon_b$ and $\epsilon_c$, thus we have utilized $NDF^\prime = d - (p+2) = d-p-2 $. This resulted in a slight, but physically not important over-estimation of the significance of the Odderon-signal when we utilized only the $\sqrt{s} = 1.96$ TeV D0 $p\bar p$ and
$\sqrt{s} = 2.76$ TeV TOTEM $pp$ data sets. The corrected value, using $NDF = d-p$ is given in Table ~\ref{tab:odsum}.

This correction improves the quality of the description of the experimental data in Ref.~\cite{Csorgo:2020wmw} and slightly decreases the significance of the Odderon observation: an at least 7.08 $\sigma$ model dependent Odderon observation reduces to an at least 6.3 $\sigma$ effect but still remains well above the 5 $\sigma$ discovery threshold, for the combined $\sqrt{s} = 1.96$ TeV D0 $p\bar p$ and
$\sqrt{s} = 2.76$ TeV TOTEM $pp$ data sets. However, it does not affect our very conservative estimate for the statistical significance of
Odderon signal at $\sqrt{s} = 7$ TeV: in Ref. ~\cite{Csorgo:2020wmw} we have just noted that this significance exceeds 7.08 $\sigma$.

Indeed, the significance of the Odderon signal in the $\sqrt{s} = 7$ TeV TOTEM  $pp$ data is extreme, we were not able to evaluate it at that time neither with MS Excel, nor with CERN Root, nor with Wolfram's Mathematica. At that time, Mathematica was able to give the best precision, but even 300 decimal digits were not precise enough to characterize the one minus the confidence level of Odderon signal in this dataset. With the help of the analytic approximations presented in ~\ref{sec:app1}, we now quantify that at $\sqrt{s} = 7$ TeV the statistical significance of the Odderon signal is greater than 37.75 $\sigma$, corresponding to a probability that starts with more than 310 digits of 9 after starting with zero and a decimal dot.
This probability in any practical terms is unity.

Table~\ref{tab:odsum} summarises all the Odderon signal observation significances in our ReBB model analysis, with the corrected application of the PHENIX method~\cite{Adare:2008cg}. The rows summarize the significances in individual datasets, measured at  $\sqrt{s} = 1.96$, $2.76$, $7.0$ and $8.0$ TeV. It is clear that the dataset at 7 TeV carries the largest, dominant Odderon signal, greater than 37.75 $\sigma$. The existence of a significant Odderon signal is confirmed with the new TOTEM data at 8 TeV, that provide an also clear-cut, greater than 18.28 $\sigma$ Odderon signal. The Odderon signal in the $\sqrt{s} = 2.76$ TeV TOTEM data is somewhat decreased, as compared to the previously published value of
7.1 $\sigma$ to a significance of $6.8$ $\sigma$, but remains safely above the discovery threshold of 5 $\sigma$.  

Now let us try to combine these statistical significances.  
These results are summarized in Table~\ref{tab:odsum-combo} and indicate the combined significances obtained with two different methods.
Given that the datasets are independent measurements, we can evaluate their combined significances step by step, by adding 
the individual $\chi^2$ and the individual $NDF$ values. These results are shown in the fifth column of Table~\ref{tab:odsum-combo}.
The sixth column in Table \ref{tab:odsum-combo} shows the combined significances by Stouffer's method ($i.e.$ by summing the significances and dividing the sum by the square-root of the number of summed significances) as used by TOTEM in Ref.~\cite{TOTEM:2020zzr}.  Stouffer's method gives a somewhat lower, more conservative estimate for the overall significance, 6.3 $\sigma$, for the combination of the results at the two lowest energies, $i.e.$ 1.96 and 2.76 TeV. However, if we also consider and combine the Odderon significanes obtained at higher energies, $i.e.$, at 7 and 8 TeV, then the statistical significance of the Odderon observation rises well above 10 $\sigma$ level, and it seems to be totally dominated by the statistical significance of the dataset at $\sqrt{s} = 7$ TeV. This suggests, that adding the 8 TeV TOTEM dataset to the sample confirms the Odderon signal, observed clearly already at lower energies.

\begin{table*}[!hbt]
    \centering
    \begin{tabular}{cccccc}
        $\sqrt{s}$ (TeV) & $\chi^2$ & $NDF$ & CL & significance ($\sigma$)   \\ \hline 
        1.96 & 24.283 & 14 & 0.0423 & 2.0  \\ 
        2.76 & 100.347 & 22 & 5.6093 $\times10^{-12}$ & 6.8  \\ 
        7 & 2811.46 & 58 & $\textless$ 7.2853$\times 10^{-312} $ & $\textgreater$37.7  \\ 
        8 & 426.553 & 25 & 1.1111$\times 10^{-74}$ & $\geq$18.2   \\ 
    \end{tabular}
    \caption{Summary on Odderon signal observation significances in the ReBB model analysis. The significances higher than 8 $\sigma$ were calculated by utilizing an analytical approximation schema, detailed in \ref{sec:app1}.}\label{tab:odsum}
\end{table*}

\begin{table*}[!hbt]
    \centering
    \begin{tabular}{cccccc}
        $\sqrt{s}$  of combined data (TeV) & $\chi^2$ & $NDF$ & CL & \shortstack{combined significance ($\sigma$)\\ $\chi^2$/NDF method} & \shortstack{ combined significance ($\sigma$)\\ Stouffer's method}  \\ \hline 
        1.96 \& 2.76 & 124.63 & 36 & 1.0688$\times$ $10^{-11}$ & 6.7 & 6.3  \\ 
        1.96 \& 2.76 \& 7 & 2936.09 & 94 & $\textless$ 9.1328$\times$ $10^{-312}$ & $\textgreater$37.7 & $\textgreater$26.9   \\ 
        1.96 \& 2.76 \& 8 & 551.183 & 61 & 4.6307$\times$ $10^{-80}$ & $\textgreater$18.9 & $\textgreater$15.7   \\ 
        1.96 \& 2.76 \& 7 \& 8 & 3362.64 & 119 & $\textless$8.0654$\times$ $10^{-312}$ & $\textgreater$37.7 & $\textgreater$32.4   \\ 
    \end{tabular}
    \caption{Summary on combined Odderon signal observation significances in the ReBB model analysis. The significances higher than 8 $\sigma$ were calculated by utilizing an analytical approximation detailed in \ref{sec:app1}.}\label{tab:odsum-combo}
\end{table*}

We investigated the case when the fit parameters of the ReBB model was optimized to the 8 TeV data. In this case the CL of the description of the data by the ReBB model improves to 14.95 \%. The extrapolated and the fitted values are the same for the $R_q$ and $R_{qd}$ parameters within triple errors while for the $R_d$ and $\alpha$ parameters within errors. Since the model parameters are correlated, the Odderon significance grows at 8 TeV when we calculate the $p\bar p$ prediction of the ReBB model by taking the radius parameter values optimized to the $pp$ 8 TeV data and the $\alpha$ parameter value from the energy dependence trend of $\alpha^{p\bar p}$. This way Table \ref{tab:odsum} contains the more conservative case when all the model parameters are taken from their energy dependence trends determined in Ref.~\cite{Csorgo:2020wmw}.


As already discussed, the $H(x)$ scaling version of the ReBB model works well in $-t$ only up to $0.97$ GeV$^2$ when the reference energy is chosen to be $\sqrt{s_0}=7$ TeV. We checked also that the $H(x)$ scaling version of the ReBB model gives a statistically acceptable description for the 8 TeV data set up to $-t=1.2$ GeV$^2$ if we assume that this dataset has also an 4.5\% overall normalization error, however, such an overall (type C) normalization error of the $8$ TeV TOTEM data set was not determined separately in Ref.~\cite{TOTEM:2021imi}, only a combination of type B (point-to-point dependent, but overall correlated) and type C (overall correlated, point-to-point independent) error was published. Note also that the 7 TeV and 8 TeV experimental $H(x)$ scaling functions are being compared in a separate manuscript, to be submitted for a publication separately,
to extend the results of Ref.~\cite{Csorgo:2019ewn} to the new TOTEM dataset measured at $\sqrt{s} = 8 $ TeV center of mass energy.
Here we emphasize that the comparison of the $H(x)$ scaling functions is a direct data-to-data comparison involving the $d\sigma/dt$, $B_0$ and $\sigma_{el}$ measurements and their uncertainties (except the type C, normalisation error, which cancels out from the scaling function). In contrast here we compared a theoretical curve to the $d\sigma/dt$ data where there are less components that are measured with uncertainties and consequently it resulted in a higher $\chi^2$ value $i.e.$ an agreement within a more limited $-t$ range.

We have also evaluated and combined the Odderon significanes obtained at higher energies, $i.e.$, at 7 and 8 TeV in this manuscript, improving on our earlier evaluation methods with analytic approximations detailed in ~\ref{sec:app1} and by estimating the number of degrees of freedom precisely in the same way as detailed in Appendix A of Ref.~\cite{Adare:2008cg}. We find that the Odderon signal is the largest in the 7 TeV TOTEM dataset, and it totally dominates even the combined signifiances.  The combined statistical significance of the Odderon observation is essentially determined by the significance at  7 TeV.  This suggests, that adding the new $\sqrt{s} = 8$ TeV TOTEM dataset to the set of analyzed data sample confirms the Odderon signal, observed clearly already at lower LHC energies, but the signal is already so clear at 7 TeV, that the observed significance of Odderon exchange within this model-dependent calculation is not increased any further. The signal of Odderon exchange cannot increase further, because  the probability of Odderon observation within this model and in the detailed kinematic range is in any practical terms is already unity.

Let us also discuss here  the already mentioned, and  currently open, ongoing scientific debate about the possibility of an Odderon contribution at $\sqrt{s} = 13$ TeV and $-t\approx0$ GeV$^2$. In Refs.~\cite{Donnachie:2022aiq,Donnachie:2019ciz} it is observed, that the $pp$ differential cross section data measured by TOTEM at 13 TeV in the Coulomb-nuclear interference region ($-t \lesssim 0.003$ GeV$^2$) and the resulting real to imaginary part ratio of the strong forward elastic scattering amplitude at $\sqrt{s} = 13$  TeV can be described without assuming any Odderon contribution at $t=0$. Hence at least one model dependent example has been given, that suggests no Odderon signal at $\sqrt{s} = 13$  TeV and at vanishing four-momentum transfers, $t=0$.
This observation, however important, does not negate our results obtained within  the ReBB model analysis presented here as well 
as published in Refs.~\cite{Csorgo:2020wmw}, since the ReBB model is calibrated and validated in a different kinematic region, in the $0.37\leq-t\leq1.2$ GeV$^2$ and $0.546 \leq \sqrt{s} \leq 8$ TeV range,
away from the optical point and lower range of center of mass energies. Thus the domain of validity of the ReBB model is clearly well outside the region where the Odderon signal may vanish according to Donnachie and Landshoff ~\cite{Donnachie:2022aiq,Donnachie:2019ciz}.

Similarly, our analysis of the $H(x)$ scaling function in Ref.~\cite{Csorgo:2019ewn} is a  direct data-to-data comparison method and not affected: 
the domain of validity of the $H(x,s|pp) = H(x, s_0|pp)$ scaling of elastic $pp$ collisions stops below $\sqrt{s} = 13 $ TeV, and all the $t=0$ observables are scaled out from $H(x,s)$. Hence the at least 6.26 $\sigma$ Odderon signal of Ref.~\cite{Csorgo:2019ewn} is also insensitive to the inspiring criticism of Refs. ~\cite{Donnachie:2022aiq,Donnachie:2019ciz}.

The value 0.14 obtained for $\rho_0$ in Ref.~\cite{Donnachie:2019ciz} does not encourage the belief that there is an Odderon contribution at $\sqrt{s} = 13$ TeV and at $t = 0$. However, there are  good theoretical reasons to believe that there is an Odderon contribution at large $-t$ and that it is identified with triple-gluon exchange. Indeed, Donnachie and Landshoff predicted \cite{Donnachie:1983ff} that $pp$ and $p\bar p$ scattering would be different around the diffractive minimum-maximum, and an Odderon contribution increasing with $-t$ was also found by using the Phillips-Barger model \cite{Ster:2015esa}. Indeed, an indication for such an Odderon signal was found with a statistical significance of 3.35 $\sigma$ already at $\sqrt{s}
= 53$ GeV at CERN ISR~\cite{Breakstone:1985pe},
an at least  3.4 $\sigma$ signal was seen in the diffractive minimum-maximum region at $\sqrt{s} = 1.96 $ TeV by D0 and TOTEM
by comparing extrapolated $pp$ data with measured $p\bar p$
data in Ref.~\cite{TOTEM:2020zzr} and a greater than 6.26 $\sigma$ signal has been seen in the comparison of $pp$ and $p\bar p$ $H(x,s) $ scaling functions in Ref.~\cite{Csorgo:2019ewn} and an at least 
6.3 $\sigma$ significance is seen in the ReBB model \cite{Csorgo:2020wmw} analysis.
In the present study we extend this latter analysis to include the final, published TOTEM results for the differential cross-section of elastic $pp$ scattering at 8 TeV~\cite{TOTEM:2021imi}.

Thus we have shown that at least within the framework of the ReBB model, and in a limited kinematic range of
$0.37\leq-t\leq1.2$ GeV$^2$ and $0.546 \leq \sqrt{s} \leq 8$ TeV, the existence of the Odderon is beyond reasonable doubt: the statistical analysis results significances much greater than 10, even 15 $\sigma$-s. Further improvements and extensions of this model are desired to be able to validate and extrapolate this model  to low four-momentum transfers, close to the kinematic limits of the optical point and to the currently largest energies available at LHC, namely to $\sqrt{s} = 13$ and to 14 TeV, the expected  largest future energies of LHC.

\section{Summary}\label{sec:summ}

As an extension to our earlier paper of Ref.~\cite{Csorgo:2020wmw} we have investigated, if the Real Extended Bialas-Bzdak (ReBB) model of Refs.~\cite{Csorgo:2020wmw,Nemes:2015iia} can describe the recently published elastic  $pp$ scattering data at $\sqrt{s} =$ 8 TeV, as measured by the TOTEM Collaboration ~\cite{TOTEM:2021imi} in a statistically acceptable way. We have also investigated if this ReBB model can be used to extract a significant Odderon signal at 8 TeV. We find that the ReBB model, at 8 TeV, works in a statistically acceptable way in the validity range in $-t$ of the ReBB model,
$0.37\leq-t\leq1.2$ GeV$^2$ and can be used to extract an Odderon signal with an extremely high statistical significance, 
as summarized in Table~\ref{tab:odsum}.  

We have also observed, that  the $H(x)$ scaling version of the ReBB model works up to $\sqrt{s} = 8$ TeV, but within  a more limited $-t$ range, $0.37\leq-t\leq0.97$ GeV$^2$. This suggests that with increasing energies, the domain of validity of the $H(x)$ scaling disappears gradually, similarly how it disappears with decreasing energies. 

\textit{Acknowledgments.} We gratefully acknowledge inspiring  discussions with A. Białas, W. Broniowksi, L. Jenkovszky, A. Kohara, T. Nov\'ak, R. Pasechnik, and A. Ster. Our research has been  supported by the NKFIH Grants no. FK-123842, FK-123959 and  K133046,
as well as by the ÚNKP-21-3 New National Excellence Program of the Hungarian Ministry for Innovation and
Technology from the National Research, Development and Innovation Fund.

\appendix

\section{An analytical approximation for significance calculation}\label{sec:app1}

The Gaussian probability density function with mean $x_0$ and variance $\sigma^2$ is given as:
\begin{equation}
    \rho(x) = \frac{1}{\sqrt{2\pi\sigma^2}}{\rm e}^{-\frac{(x-x_0)^2}{2\sigma^2}},
\end{equation}
with
\begin{equation}
    \int_{-\infty}^{\infty}dx\rho(x)=1.
\end{equation}

The confidence level (CL) is then calculated as:
\begin{equation}
    {\rm CL} = \frac{2}{\sqrt{2\pi\sigma^2}}\int_{x_0+\delta}^{\infty}{\rm e}^{-\frac{(x-x_0)^2}{2\sigma^2}}dx,
\end{equation}
where $\delta=n\sigma$. Applying a variable change, $$x\rightarrow x' = x-x_0-\delta,$$ we have
\begin{align}
    {\rm CL} = \frac{2}{\sqrt{2\pi\sigma^2}}\int_{0}^{\infty}{\rm e}^{-\frac{(x'+\delta)^2}{2\sigma^2}}dx' \\ \nonumber \leq \frac{2}{\sqrt{2\pi\sigma^2}}{\rm e}^{-\frac{\delta^2}{2\sigma^2}}\int_{0}^{\infty}{\rm e}^{-\frac{x'\delta}{\sigma^2}}dx' = \sqrt{\frac{2}{\pi}}\frac{\sigma}{\delta}{\rm e}^{-\frac{\delta^2}{2\sigma^2}}.
\end{align}
Thus finally one obtains:
\begin{equation}
CL \leq \sqrt{\frac{2}{\pi n^2}}{\rm e}^{-\frac{n^2}{2}} .
\end{equation}
This formula gives the lower limit for the significance $n$ in $\sigma$-s corresponding to a given CL value. This formula is useful if numerical calculations fail when the value of CL is too small as in our case.

\bibliographystyle{spphys}  
\bibliography{Odderon-Letter}

\end{document}